\documentclass[final,numberedheadings] {aipproc}

\layoutstyle{8x11double}

\begin{document}
\def\OMIT#1 {{}}
\newcommand{\beq}{\begin{equation}}
\newcommand{\eeq}{\end{equation}}
\newcommand{\eqr}[1]{(\ref{#1})}
\newcommand{\half}{\frac{1}{2}} 
\newcommand{\la}{{\langle}}
\newcommand{\ra}{{\rangle}}
\newcommand {\HH} {\mathcal{H}}  
\newcommand {\ep}{\epsilon}
\newcommand {\epdot} {\dot{\ep}}

\newcommand{\xaxis}{{\bf \hat{x}}}
\newcommand{\yaxis}{{\bf \hat{y}}}
\newcommand{\zaxis}{{\bf \hat{z}}}
\newcommand{\rr}{{\bf r}}
\newcommand{\EE}{{\bf E}}
\newcommand{\rcargo}{{r_{\rm cargo}}}

\title{Possible mechanisms for initiating 
macroscopic left-right asymmetry
in developing organisms}


\author{Christopher L. Henley}{
  address=
{Department of Physics, 
Cornell University, Ithaca, New York 14853-2501, USA}
}

\begin{abstract}
How might systematic left-right (L/R) asymmetry of the body plan 
originate in multicellular animals (and plants)?  
Somehow, the microscopic handedness of
biological molecules must be brought up to macroscopic scales.
Basic symmetry principles suggest that the usual ``biological'' mechanisms 
-- diffusion and gene regulation -- are insufficient to implement the
``right-hand rule'' defining a third body axis from the other two.  
Instead, on the cellular level, ``physical'' mechanisms (forces and
collective dynamic states) are needed involving the long stiff fibers
of the cytoskeleton.
I discuss some possible scenarios;
only in the case of vertebrate internal organs
is the answer currently known (and even that is in dispute).
\end{abstract}

\maketitle


\section{Introduction}
\label{sec:intro}

The anatomy of most animals -- and many plants --  
breaks left/right symmetry,  the same way for all
or most individuals (though this doesn't
necessarily have any functional significance).
The key processes at the cell or organism level --
diffusion, regulation of gene expression, perhaps
elasticity -- don't distinguish left from right,
so how can the developing organism ``learn'' this
\cite{wolpert-brown, wood-review,levin-LR-review,wood-review-2005}~?

Of course, the proteins (or other biological molecules) that
constitute the cells are handed;~\footnote{
It should be emphasized I am {\it not} interested
here in the original prebiotic symmetry-breaking which 
determined the molecules' 
handedness~\cite{frank53,avertisov-molecules,garcia-bellido}.}
but how can this information be brought to the macroscopic scale?
Any mechanism for that, I suggest, involves {\it physics}
to an unusual extent: forces and motions acting on the stiff
semi-macroscopic polymers that make up each cell's ``cytoskeleton''.
Furthermore, I argue that just from {\it a priori}, basic 
{\it symmetries}, the possible mechanisms are strongly constrained.
That is why this topic seemed appropriate for a conference in
honor of Landau, even though he would probably have considered
the entirety of biological physics to be ``pathological''.

One motivation to pursue L/R specification is ``cleaner''~\cite{palmer}  
than most development instabilities in that 
(i) it emerges out of a functionally symmetric state (if not,
the original L/R specification just was earlier). Maybe, e.g., the
differentiation of brain regions emerges in a similar fashion, but
the way it goes can be explained by the morphological asymmetries of 
neighboring tissues;
(ii) it is binary -- the minimum of information;
(iii) mutant embryos have unambiguous visual signatures.
Perhaps for these reasons, L/R experiments have surged 
over the past decade.

The rest of the paper begins with an introduction in 
which I list examples of L/R asymmetry in model organisms, 
lay down key facts and assumptions, and
classify the mechanism.  The next four sections go on to 
tell four stories of L/R asymmetry (of which only the first
is experimentally settled): 
cilia driving fluid flow in vertebrates (Sec.~\ref{sec:nodal-flow}),
a hypothetical mechanism wherein screw
processive molecular motors transport signaling molecules
(Sec.~\ref{sec:screw-transport}),
shearing actin arrays causing a twist in cell division as in molluscs
(Sec.~\ref{sec:celldiv}), 
and finally rotating microtubule arrays in 
{\it plants} leading to a macroscopic twining as in vines
(Sec.~\ref{sec:mt-plants}).

\subsection{Examples}

The first example is the internal organs of 
vertebrates~\cite{cooke-review,hirokawa-review}: 
heart, lungs, etc. are located asymmetrically.
In humans,~\cite{mcmanus-RHLH}
the frequency of mirror-reversal is $\sim~10^{-4}$.
Common model species are  mouse, chick, Xenopus frog,
or zebrafish.

A second example is the human brain:~\cite{sun-brain-review}:
of course, right-hand
dominance is a side effect of left-brain dominance.
The congenital reversal frequency~\cite{mcmanus-RHLH}
is $\sim 10^{-1}$, and
surprisingly is {\it independent} of the handedness
of internal organs~\footnote{
See references in \cite{mcmanus-RHLH}, pp. 5-6; \cite{cooke-review},
Part IV; and \cite{levin-LR-review}, Sec. 10.
This independence is also confirmed in frogs~\cite{malashichev-review}.}
so it probably has a different mechanism.
(While some anatomical and functional brain asymmetries are 
found in other animals, the relation to human brain 
asymmetry~\cite{malashichev-review,halpern-LR,malashichev-book}
is still unclear.)

Thirdly, even the humble {\it C. elegans} nematode ``worm'' is 
L/R asymmetric
~\cite{wood-review,wood-review-2005,wood-orig,poole-hobert,hobert-brain-review}
--- a creature so small that all $10^3$ cells are numbered by
embryologists (they develop in an identical, and known, pattern),
and every embryo repeats exactly the same sequence of divisions.
Here, the gut twists, and a certain chemosensing
neuron has functional L/R asymmetry.
(Flies also have twists in their guts and genitals~\cite{
speder-drosophila-orig,hozumi-drosophila-orig,speder-noselli-review0}.)

As a fourth and final animal example, 
mollusc shells --- say snails --- are well known to coil 
right-handed (reversal frequency $\sim~10^{-4}$).
This has functional consequences in that mating is
awkward between snails of different
handedness,~\cite{snail-species} and 
a snake has evolved asymmetric jaws~\cite{snail-snakes}
to better crush right-handed snails.

There is a parallel story of handedness in plants 
(Sec.~\ref{sec:mt-plants})
--- certainly in climbing plants, whose roots and shoots both spiral with a 
species-dependent handedness.~\footnote{
Note that ``phyllotactic spirals'' in plants, responsible for
Fibonacci numbers in leaf placement, in pine cones and sunfloweers --- 
do {\it not} have a fixed handedness.}
(An interesting symmetry corollary
--- this would be obvious to Landau! --- is that roots growing 
against a vertical hard surface can deviate by a ``Hall angle'' 
from heading straight downwards, and they do.)

Since this is biology, we do not expect universal answers;
if a conjectured mechanism turns out to be wrong for one
mentioned example, it might be valid for another!




\subsection{Question and starting assumptions}
\label{sec:question-assumptions}

An embryo can develop two axes by spontaneous symmetry breaking:
(i) the {\it anterior/posterior} (A/P) axis, i.e. head/tail ---  call this
the $\xaxis$ axis; (ii) the
{\it dorsal/ventral} (D/V)  axis,i.e. back/front --- call this
the $\zaxis$ axis.
It is well understood that a combination of reaction and diffusion by
chemical signals can generate this sort of pattern formation. (Here
``reaction'' includes regulation of DNA transcription and translation
to proteins.) Note that in practice  the symmetry breaking is 
often biased externally, e.g. by the point sperm entered egg.

A third ($\yaxis$) axis could certainly form, normal to the others, 
by a spontaneous symmetry breaking.
But the ``right-hand rule'' 
   \begin{equation}
     \yaxis = \zaxis \times \xaxis
   \label{eq:rhrule}
   \end{equation}
must be ensured: how can it be done?
The key claim of this paper is that symmetry requires that 
{\it each of the three symbols} on  the right-hand side of 
\eqr{eq:rhrule} has {\it a specific physical (biological) correlate} 
entering the mechanism.
That is, there must be two kinds of preexisting polarization
representing the $\xaxis$ and $\zaxis$, plus some functionally
chiral element representing the ``$\times$'' in \eqr{eq:rhrule}.

I mention two important assumptions.
First, there is a {\it spontaneous} symmetry breaking:
some robust mechanism ensures an asymmetric outcome,
but that by itself would produce an equal mixture of L and R organisms.
(This is evidenced experimentally, in some cases, in mutants that
have such a randomization.)
The uniform outcome is due to an additional {\it small} biasing field
(exponentially small in the system size),
just as a tiny magnetic field 
decides the magnetization sense of an Ising magnet
cooled through its Curie temperature. 
The practical significance is that our mechanism need only produce a 
weak bias (let's arbitrarily aim for $\sim 10^{-2}$), 
since the assumed symmetry breaking amplifies it enormously.~\footnote{
Symmetry breaking may also explain the functional reason for a bias, when
there is no social reason for organisms to all have the same handedness. 
In the absence of a bias field, occasionally an organism would be a mixture of
``left'' and ``right'' type domains, which would be a congenital defect.}

Second, I also assume that L/R asymmetry stems from
the microscopic chirality of molecules (under genetic control),
and {\it not} e.g. an asymmetry produced in the egg by
a right-handed mother.  (This is easily refuted by the inheritance patterns;
and wouldn't such a  mechanism have $>10^{-4}$ error rate?)

Finally, since we seek the {\it earliest} L/R asymmetry,  it follows (tautologically)
that whatever caused it must have been L/R symmetric.~\footnote{%
Since all the molecules are asymmetric, more carefully we should say the cause is 
{\it functionally} L/R symmetric; just what that means will be clarified by
examining the specific scenarios worked out in later sections.}

\subsection{Classifying mechanisms}
\label{sec:classify}

There are three useful categories.

\subsubsection{Two levels of any mechanism}

Any explanation of L/R specification really requires two stories, 
one at the cellular level and one at the collective level.
The cell level story starts from proteins --- that's where the cytoskeleton 
fibers come in --- and goes to properties of the whole cell, or the
interaction of one cell and a neighbor.  
The collective story starts from the cell behavior and explains how
this specifies a left and a right side in the whole embryo.
For each of the following sections, I will indicate both levels;
sometimes one or the other is rather trivial, but there are always
the two levels.  The symmetry principle applies at {\it each} level.

\subsubsection{Two styles in development}

In embryology,~\cite{wolpert-text}
there are two general ``styles'', applying to different
animal phyla.
An ``early'' style applies to molluscs and {\it C. elegans}:
as cells first divide, each gets a determining label, schematically
like a binary string. All its descendents retain that string, while 
possibly adding bits that refine the specification of cell type 
in the mature animal.  Thus, cell fates are fixed early.
On the other hand, vertebrates and insects have a ``late''
style: through many cell divisions the cells are unspecified,
then fates are undetermined by pattern formation within
a multicellular embryo.
Evidently, an  essentially cell-level L/R mechanism suffices for early-style
creatures, whereas a collective L/R mechanism seems to be needed in late-style
creatures. 

\subsubsection{Two ways to represent L and R}
\label{sec:ways-represent-LR}

\begin{figure}
\includegraphics[width=0.90\linewidth] {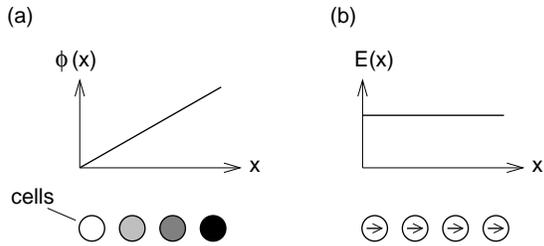}
\caption{Two ways to represent left/right information
(a). Positional information (concentration)
(b). Polarization (gradient)}
\label{fig:LR-represent}
\end{figure}

There are two ways that ``leftness'' might be represented in an embryo
(Fig.~\ref{fig:LR-represent}).
One is called ``positional information'':~\cite{wolpert-text}
it means some chemical has
a concentration $\phi(x,y,z)$ which is roughly a function of (say) $y$;
three such chemicals can specify all coordinates.  By sensing the
concentrations of all three, a cell learns its position within the body
and hence which organ it should become.  (To do that, it does not necessarily
need to know which {\it direction} is left, or posterior, etc.)

The alternate representation is ``polarization'' of cells, so that 
each ``knows'' which {\it direction} is left, but not where it sits
along the L/R axis.   We could write this as a local vector $\EE(x,y,z)$.
Evidently the relation of polarization and positional information is
$\EE = \nabla \phi$.  
Wolpert's pioneering paper~\cite{wolpert-brown}
about the L/R mechanism as a symmetry problem 
envisaged a ``polarization'' representation.  
I believe that was meant only as a thought experiment,
rather than a literal proposal for the mechanism;
``positional information'' seems in many cases easier 
to generate,  as well as being the information 
ultimately needed to ``inform'' a developing tissue 
as to its fate.

It is not trivial to convert one kind of L/R information to the other.
Indeed, given $\phi(\rr)$, a eukaryotic cell is big enough and sophisticated
enough to sense a concentration gradient between one side
and the other, and develop a polarizations in response, to 
differentiate (in the calculus sense!).  But the reverse construction --- 
integratiing --- cannot be done locally.  To generate an imbalance in 
the concentration, some chemical it must get actively transported through
the organism with a bias along $\EE (\rr)$.  

\subsection{Cartoon of the cytoskeleton}

The mechanism is not of the usual ``biological'' type.
That would mean transport 
(by diffusion or otherwise) of signaling molecules and reactions.  
But transport alone won't suffice for L/R, since diffusion 
doesn't distinguish handedness.
All possible mechanisms seem to involve actual 
{\it forces} or {\it torques} 
exerted by molecular motors, which somehow structure the 
{\it cytoskeleton} made of stiff, semi-macroscopic fibers
(that the motors run along).

\OMIT{Transport creates L/R bias in concentrations only if
   the network geometry already had a bias}

The cytoskeleton is the framework in each cell of a eukaryotic (higher) organism,
built from long, stiff, directed, and helical macro-molecules (which I will
call ``fibers''); there are specific kinds of motor molecules 
for each kind of fiber.~\cite{cytoskeleton}.
The two kinds of fiber are (i) {\it microtubules} (mt), with {\it dynein} 
or {\it kinesin} 
motors on them;  (ii) {\it actin} fibers,   with {\it myosin} motors on them.
Each family of motors contains numerous subvarieties, used by the cell for
special purposes. In particular, myosin V is the main myosin variety that moves along 
a fiber for long distances (``is processive''); other varieties, e.g. myosin II
used in muscles, only makes contractions.
We note that, despite the microscopic differences between mt and actin,
when abstracted to the model level they may look very similar: i.e., 
a physics approach may uncover (even quantify) a kind of universality.  

\begin{figure}
\includegraphics[width=0.75\linewidth]{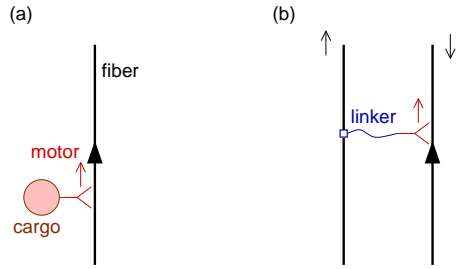}
\caption{Two-legged molecular motors walk on fibers
(a) carrying cargoes (b) linked to other fibers and driving
a relative sliding.}
\label{fig:motors}
\end{figure}

The two directions along the fibers are not symmetry-equivalent -- the
growth direction is called the fiber's ``polarity''.
A given kind of motor (literally) walks in a fixed sense
(which we assume to be the $+$ sense for this paper).
It transports ``cargoes'', which  are typically chemicals in 
vesicles (small membrane bags attached to the motor by linking
protein(s): see Fig.~\ref{fig:motors}(a)).
Occasionally (by thermal fluctuation) a
motor falls off its fiber, and diffuses till it reattaches to 
the same or another fiber

In place of a cargo, a motor can be linked somehow
to another fiber [see Fig.~\ref{fig:motors}(b)] 
so as to drive the motion of one relative to the other
{\it e.g.} in cell division.
Fibers form networks with many crosslinks, which 
are often {\it dynamic}, i.e. 
fibers are constantly appearing, growing, shrinking, and vanishing,
producing a dynamic steady state describable by statistical mechanics.

\section{Nodal flow mechanism}
\label{sec:nodal-flow}

Here I merely review the only well-understood mechanism, 
a {\it late-stage} type mechanism acting in 
{\it vertebrates}~\cite{hirokawa-review}.

\subsection{Cell-level story}

We start with an approximately flat embryo that has already 
formed A/P and D/V axes.  
The following sentence describes the key and sufficient cause:
watch carefully to see where each element of
\eqr{eq:rhrule} gets mentioned.
On its ventral side, near the ``node''  (a key place in a 
developing vertebrate embryo), are special cells with {\it cilia} ---
moving tails that stick {\it out} (``$\zaxis$'') from surface, but
{\it tilted} (``$\xaxis$'') 
towards the embryo's  posterior end.~\cite{PNAS-nodecilia}.
These cilia move {\it circularly} ($\times$), 
--- clockwise, looking down --- unlike regular cilia 
(which move back and forth).

\subsubsection{Root of L/R asymmetry}

So, {\it why} do the node cilia move circularly?  
This is plausible from their structure:
each cilium has 9 pairs of microtubules that run its length, 
and each pair is linked by  dynein molecules with their
head-to-foot direction oriented {\it clockwise}.
The crucial L/R event is that a ring of special protein 
molecules assembles in the cell membrane. This serves as
the template to start the microtubule pairs with dynein
linkages, and the whole structure apparently grows outward
ring after ring by stacking each component onto its
own kind in the previous ring.
In the end, then, the L/R asymmetry is not due to the
inherent helicity of the microtubules, but rather the
handedness of the templating complex.
Plenty of physics remains to be worked out in this system, namely 
applying elastic theory and fluid dynamics to show why 
the given structure executes circularly polarized motion, 
and in which sense~\cite{brokaw-chiral}.

\subsection{Collective level (and alternate story)}

The array of cilia just described is sufficient to break L/R symmetry:
symmetry allows it to drive a fluid flow L to R across the embryo
(as observed).  If a signaling chemical is  released, the
flow carries it to the L side,  where it can bias the symmetry breaking.
The key check is that you reverse the flow externally  and get 
out reversed embryos --- in {\it mice}.  

But the relative importance of 
early versus late mechanisms may depend on the kind of vertebrate:
early L/R asymmetries were clearly seen in 
{\it Xenopus} frogs~\cite{levin-ionflow-review},
due to a distinct L/R mechanism.
Possibly before the first cell division, preexisting
(maternal) chemicals -- in particular ion transporter proteins
-- are getting asymmetrically distributed in the egg;
plausibly the L/R asymmetry comes via some 
kind of actin/myosin mechanism like the one 
discussed in Sec.~\ref{sec:celldiv}, below~\cite{levin-aw}.
The collective level of this mechanism involves a biased
transport that converts an electrical potential difference
into a concentration by transport (like the ``integration''
process of Sec.~\ref{sec:ways-represent-LR}).
It was speculated the collective mechanism includes a
mutual feedback process (the spontaneous symmetry breaking
assumed in Sec.~\ref{sec:question-assumptions}),
between the electric field and serotonin concentration
gradient~\cite{levin-serotonin}

\OMIT{If the cilia stuck straight out, we still can get 
  fundamentally same symmetry breaking.  (Now flow
is circular, and signal must be released from, say,
 the head end.)}

\section{Asymmetric transport}
\label{sec:screw-transport}

This section outlines, as a pedagogical example, 
a completely hypothetical mechanism;   it is late-stage type
and (unlike the mechanisms of Sections \ref{sec:celldiv}
and \ref{sec:mt-plants})
it depends on transport of signaling molecules, like
typical ``biological'' mechanisms do.
The key ingredient is helical motion of motor molecules on
a cytoskeletal fiber.

Consider for simplicity an embryo  with its geometry 
flattened into the plane the normal to the D/V axis.
Assume the dorsal and ventral sides are distinguished
(there's the ``$\zaxis$ again, from \eqr{eq:rhrule}).
We look for an analog of  the Hall effect whereby (in the
presence of a transverse magnetic field) electrons drift
at a small angle from the electric field direction.
Here, chemicals transported due to an assumed A/P polarization 
(``$\xaxis$'')
actually move at an  angle rotated slightly (``$\times$'') 
from the A/P axis, allowing them to get carried preferentially 
to one side of the embryo and thereby bias the assumed symmetry breaking.
The upshot will be that it's not very easy to engineer 
late-stage asymmetry with a transport mechanism!

\subsection{Cell level mechanism}

This is the nontrivial level for this mechanism; again,
I'll tag the three ingredients from \eqr{eq:rhrule}.
Let's assume a ``cortical'' array of fibers, meaning 
it is just {\it under} the cell's membrane (there's the ``$\zaxis$''). 
Let the fibers be
oriented along $\pm \xaxis$; no net polarization is assumed, 
so we are {\it not} yet talking about the the A/P 
($\xaxis$) asymmetry.

\begin{figure}
\includegraphics[width=1.00\linewidth]{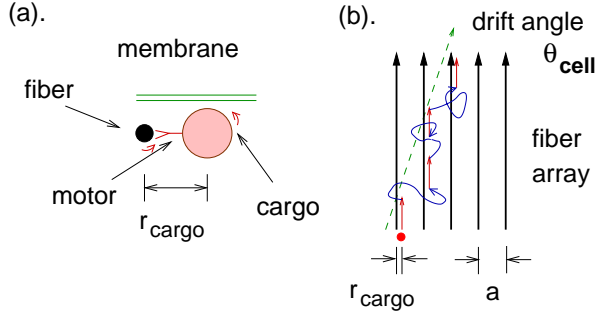}
\caption{Mechanism of transverse drift of a ``cargo''
due to a spiraling processive motor.
(a) End view (the motor is moving into the page
(b) Top view (membrane is the plane of the page);
the mean direction of transport is tilted by an
angle $\theta_{\rm cell}$.
}
\label{fig:cargodrift}
\end{figure}

Active transport relies on motor molecules; since the fibers
they move on are (microscopically)  helical , their motion should
(generically) be helical too [there's the ``$\times$ in \eqr{eq:rhrule}].
In fact, myosin V motors are known to spiral 
as they move on actin~\cite{ali-helical}.
(However kinesin motors are tightly bound to the protofilaments that
constitute a microtubule and those are, most often, quite straight.)
Now, let's imagine our motor spirals {\it counter-clockwise} (ccw);
this will pull the cargo vesicle around the fiber till it gets
jammed against membrane, and from then onwards the motor will just
move lengthwise:  in other words, it always travels 
``on the right side of the road'' [see Fig.\ref{fig:cargodrift}(a,b)].

Each time the cargo detaches, diffuses, and reattaches during its progress,
the most likely outcome is to reattach to the {\it same} fiber as it is
closest.  If it reattaches to a different fiber, it's likelier to reattach 
to the fiber on its {\it right} side [Fig.\ref{fig:cargodrift}(b)] since that
one is always closer than the  one on the left.
When we model this diffusion as confined to the plane (hence basically 
{\it one} dimensional), the probability of hopping one fiber to the
right is $\rcargo/a$, where $a$ is the fiber spacing\cite{berg}; it is
still proportional to this ratio in more realistic models.
If the cargo succeeds in hopping over, its path gets shifted by $a$ 
to the right: hence, the {\it average} non-random shift is $\sim \rcargo$.

The result is a mean transport current rotated rightwards 
by $\theta$ from $\pm \zaxis$, where
       \begin{equation}
                \theta_{\rm cell} \sim \rcargo / \ell
       \label{eq:theta_cargo}
       \end{equation}
Here $\ell$ is the typical distance this motor goes 
before falling off (``processivity''), 
and $\rcargo$ is an effective distance of the cargo from 
the fiber axis.  
If we modeled the free diffusion interlude as a one-dimensional random
walk transverse to the fiber array, the probability of reattaching to 
the fiber on your right is $\rcargo/d$ where $d$ is the separation;
the likeliest outcome is to reattach to the same fiber, since it's closer.
Since $\ell \sim 1~\mu$m and $\rcargo \sim 25$ nm, we find
$\theta_{\rm cell} \sim 2.5 \times 10^{-3}$.

\subsection{Collective level}

\begin{figure}
\includegraphics[width=1.00\linewidth]{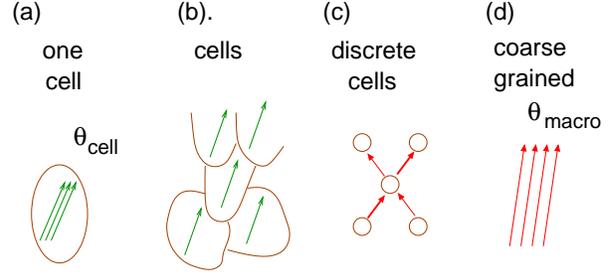}
\caption{Collective level story:
if (a) the transport in each cell is rightward-biased,
so is (b) the relative probabilities of transfer to a neighbor,
represented in (c) as a branching ratio in a discretized
model where each cell is one node of a network.  
Coarse-grained further to a continuum model, we end up
again (d) with a bias angle.
}
\label{fig:drift-coll}
\end{figure}

This array is found only on (say) the ventral side of cells
on (say) the embryo's ventral side (the ``$\zaxis$'' at organism level).
Imagine the signal chemical gets released from (say) the anterior
(head) end ({\it this} is the A/P or ``$\xaxis$'' asymmetry at last).
If the signaling chemical has a sideways bias $\theta_{\rm cell}$
relative to the array in each cell, it's easy to see the 
macroscopic transport will have a bias $\theta_{\rm macro}$ 
of similar magnitude (see Fig.\ref{fig:drift-coll}).

\section{Cell division}
\label{sec:celldiv}

I now turn to an example that is not yet understood, but
we know so much from experiments that we {\it should} be
above to pin down the mechanism.  This is 
(to use the categories of Sec.~\ref{sec:classify}) 
the early-stage, cell-level mechanism 
of twisting cell division 
(``spiral cleavage'' to the mollusc community),
apparently due to an array of actin filaments.
First I will review some experimental facts.

For {\it molluscs}, specifically snails,~\cite{kuroda-snail-actin}
after the first two divisions an embryo consists
of four cells in a square.
These give rise to four (smaller) daughter cells by
dividing along the axis normal to the square, but
break the symmetry by twisting to the right till
the new cells sit on top of the furrows between the
first four cells.
This determins the handedness of the grown snail
(one evidence is that mutants which divide at this
stage with the opposite twist, 
also have macroscoically reversed handedness.)

In {\it C. elegans} embryos, too, the embryo with four
cells (planar but less symmetric) is the
stage from which the ultimate handedness is 
determined.~%
\cite{wood-review,wood-review-2005,wood-orig,poole-hobert,hobert-brain-review}.
This was proven by manipulations wherein the cells get physically 
switched, leading to a mirror-reversal in the grown animal~\cite{wood-orig}.
Furthermore, the motions of the cells in these divisions 
can be described by a general twisting tendency like
the mollusc case~\cite{wood-review-2005}.

\subsection{Organism-level story}

Putting this all together, we have a mechanism for
the organism level.  We hypothesize a torque which
always drives the same twist of two 
daughter cells about their axis of division.  
If the cells were already latently polarized along parallel
axes for the subsequent division, those axes get
tilted by the torque as observed.
Then right after the symmetry-breaking  division at the four-cell
stage, some chemical signal is passed  between the cells,
depending on which cell neighbors.
Since the neighbor relation has become asymmetric, 
this tells the cells which is L and
which is R; that ```bit'' of information is 
preserved in subsequent divisions for the descendents
of these cells, and is expressed functionally
at a much later stage.

\subsection{Microscopic level story?}

Experimentally it was shown that the twist depends on {\it actin} 
but {\it not} on microtubules,~\cite{kuroda-snail-actin}
somewhat surprisingly since
microtubules have the more prominent role in cell division
(forming the ``spindle''  between the two new cell nuclei.)
The role of actin in cell division is to form the
``contractile ring'', an array 
of roughly parallel filaments that contract to pinch off
the two cells from each other~\cite{contractile-ring,contractile-PRL}.

Meanwhile, in frog eggs (under the influence of a certain drug),
a twist of just this sort is observed and was shown to depend
on an actin array.~\cite{danilchik-xenopus-twist}.
Bundles of parallel actin fibers {\it shear} past 
each other, always in a {\it clockwise} sense. 
\OMIT{and with a fairly uniform shear}
It appears myosin is responsible, 
rather than actin polymerization, as shown by turning off the latter
with a poison.  (In the case of {\it Drosophila}, 
a {\it late}-style mechanism which might or might not be related
to this one, myosin I D is responsible~\cite{%
speder-drosophila-orig,hozumi-drosophila-orig,speder-noselli-review0}.)

\begin{figure}
\includegraphics[width=0.8\linewidth] {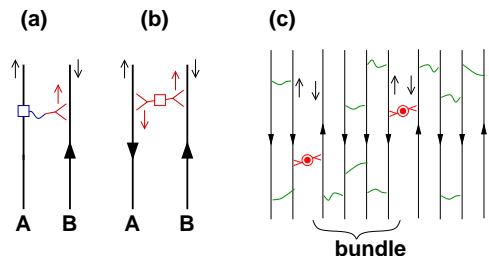}
\caption{
The relative shearing of two actin fibers demands a handedness
relationship between the sense of their polarization(s).
To drive clockwise shear, the dominant 
bridging bonds must  be placed as shown in (a,b)
sticking to the right or the left, according to the polarity of 
the walked-on fiber.   
Large arrows in the fibers indicate their polarity; in (a), 
the polarity of fiber A does not matter.
The motor-to-actin or motor-to-motor
linkages are shown as squares in (a,b); small arrows show the motion
of motor relative to fiber, or fibers relative to each other.
In (c), bundles are shown connected by static linkers that do not
restrict the fibers' polarities; a special linker shown 
as $\bigodot$, binding  two (oppositely oriented) motors to
the membrane, can drive a shear only in the clockwise sense.
}
\label{fig:array-shear}
\end{figure}

\OMIT{
{\bf Assume:} formation of parallel actin  array doesn't depend on 
LR.  Hence it's equal mixture of both polarities.  ({\sl Alternatively:}
may have an overall polarity, but either sense equally likely.)
}

I do {\it not} have a satisfactory microscopic model for this case.
The experiment shows the actin array's shear motion is driven by 
myosin, and in turn the myosin moving on one actin
fiber must be linked (directly or indirectly) to another fiber, 
in order to make any shear (see Fig.~\ref{fig:array-shear}).
So, the actin array somehow becomes organized --- either 
(i) driven by the myosin motors, or (ii) during the array's
formation --- such that whenever a motor is attached (in any fashion) 
to actin fiber $A$ and walking along actin fiber $B$
(as shown in Fig.~\ref{fig:array-shear}),
the polarity of fiber $B$ is towards the left
as seen from the $A$ -- $B$ link.~\footnote{
A special case of this is where the linkage connects
motors walking on {\it both} actin fibers 
[e.g. Fig.~\ref{fig:array-shear}(b)], as in muscle.}

This arrangement might be produced (i) conceivably, by the spiraling
of processive motors, as in the hypothetical mechanism of 
Sec.~\ref{sec:screw-transport}.  But (ii)
a more plausible mechanism would depend on the
linker proteins that bind actin fibers into bundles (Fig.~\ref{fig:array-shear}(c)
A ``basic'' linker would join fibers with the {\it same} polarity. 
But a ``special'' linker, joining fibers with the {\it opposite}
polarity, would be a dimer of {\it membrane} anchored proteins oriented 
to hold the motors oriented as shown, so they drive relative motion
only if both fibers have polarity to the {\it left} as seen from
the link.  

\OMIT{Imagine a myosin V motor bonded 
to fiber A,  and walking on fiber B.
The motor spirals around fiber B 
till it comes against membrane.
If the motor is on the {\it far} side of fiber B,
linker strained and hence myosin falls off easier.
Thus, the near-side case dominates.
(To estimate this quantitatively depends on knowing the
linker's elasticity.)}

\section{Plants: rotating mt array?}
\label{sec:mt-plants}

Finally, I turn briefly to the case of plants.
Like the previous story (cell division),
this involves a dynamic, cortical 
fiber array --- this time of microtubules (mt).


\subsection{Organism scale mechanism}

Plant cells are cylindrical and elongated in the growth direction.
An mt array~\cite{hashimoto-mt-review}
forms around their cell walls, roughly parallel and nearly
transverse to the cylinder, but with a typical helical pitch
(say a pitch angle $\theta_0$ away from transverse).
In turn the microtubules orient a helical array of cellulose fibers
around the cells, which stiffens the cell walls.

Now, the shoots (and roots) of many plant species have 
a consistent helical sense, particularly evident in 
climbing vines that twine around vertical supports.  
Experiments on {\it arabidopsis} confirmed the sense of the 
plant's macroscopic twist corresponds to that of the
microtubule array's microscopic twist on the membrane:
mutations that reverse the latter also reverse the former
~\cite{hashimoto}.

\begin{figure}
\includegraphics[width=0.4\linewidth]{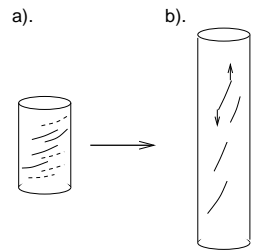}
\caption{
Origin of twist torque in a plant cell.
(a).  The cylinder represents a cell wall.
On it, cellulose microfibrils (thick segments) are templated to grow 
with a small tilt from the circumferential direction.
(b).  The cell grows by elongation. (Exaggerated in figure).
The microfibrils do not elongate, but are forced to twist, 
so each is under tension and exerts a torque.}
\label{fig:plantcell}
\end{figure}

My conjecture for how this happens
comes from a paper about fungi~\cite{gamow}, 
where chitin plays the role of cellulose.
(A related but not identical mechanism was
proposed for helical twisting in chains of
elongating bacteria without flagella~\cite{wolgemuth}.)
Imagine a cellulose fiber of finite length $L$ anchored in the membrane
(Fig.~\ref{fig:plantcell}).  
The cell grows by elongation but the fibril can't. 
Hence it feels opposing longitudinal forces at the ends, roughly
proportional to $\epdot L \cos \theta$, where $\epdot$ is the 
fraction elongation per unit time.  Since the fiber isn't
longitudinal, these forces exert a twist torque on the fiber (and vice versa)
proportional to $\epdot L \cos\theta \sin\theta$.
Thus all the individual cells get a microscopic torque stress of 
the same sign, adding up to a macroscopic one on the whole shoot.
Now elasticity theory tells us there can be an instability:
if the centerline deviates from straight to helical, this 
can relieve the torque stress and decrease the energy 
(Such a ``twist-to-writhe'' conversion.
is also responsible for supercoiling of, e.g., DNA.)

\subsection{Cell level mechanism: mt arrays}
\label{sec:plant-cell}

Microtubule arrays get oriented in a collective process with 
the following rules:~\cite{mt-rules}
\begin{itemize}
\item{Rule (i)}
The $+$ end of the mt (mostly) grows, while the $-$ end
depolymerizes (but not as fast). 
\item{Rule (ii)}
Furthermore, new mt
nucleate  on existing mt and grow at a specific branching angle
$\sim 30^\circ$.  
\item{Rule (iii)}
When a growing mt hits another, if the relative
angle is less than $\sim 30^\circ$ it bends 
(this demands a linker to exert a force, since mt are rather stiff) 
and aligns (this process forms bundles).
\item{Rule (iv)}
If the angle is larger, the growing mt suffers a ``catastrophe''
meaning it {\it depolymerizes} from the $+$ end (and disappears).
\end{itemize}
Note that either process (iii) or (iv) tend to drive the mt towards a
steady state phase in which the alignment has long range order,
as found in a simulation~\cite{mt-simulation}
that omits branching (ii) and bending (iii)].
Also, the collective state is symmetry-broken: within the 
plane of the membrane, the mt array's orientation is 
(so far) arbitrary.

It seems difficult to engineer such a mechanism to directly orient
the mt with respect to the elongation axis.  Instead, I speculate
there is a spontaneous (slow) {\it rotation} of the array.
The overall mean orientation angle $\theta(t)$ follows the
dynamics
  \beq
      \frac{d\theta}{dt} = \omega  - (\alpha-\epdot) \sin 2\theta 
  \label{eq:theta-dot}
  \eeq
Here $\omega$ is the spontaneous rotation rate (see below
for possible mechanisms).  The term proportional to $\epdot$
expresses how elongation passively carries the mt to a higher angle.
The consequence of \eqr{eq:theta-dot} is the angle will evolve
to a steady value $\theta_0$ satisfying
    \beq
       \sin 2\theta_0 =  \omega/(\alpha-\epdot).
    \eeq
To explain the small (=nearly transverse) observed $\theta_0$,
we must posit the additional transverse bias $\alpha$, which 
tries to pull $\theta$ back towards 0 or $180^\circ$.
(Here is one speculative mechanism for $\alpha$: 
say there are membrane anchoring 
proteins that slightly bend the mt away from the membrane,
so in effect the mt has a spontaneous curvature.
If so, the mt has the least strain when it aligns along the 
direction of the membrane's maximum curvature.)

Such dynamic rotation is suggested by morphologies of some
plant cells that have multiple layers of cellulose fibers,
each rotated relative to the one underneath~\cite{preston82}.
Indeed, domains undergoing such a rotation were seen directly 
by video imaging in growing plant cells~\cite{mt-video}
(under the influence of a drug?).
The authors do not infer a particular sense, but about 70\%
of the domains rotated clockwise (see Fig. 1 of Ref.~\cite{mt-video}).

So the $\omega$ term in \eqr{eq:theta-dot} is the key
parameter determining L/R; what  are the mechanisms for it?
Recall that, by our basic symmetry arguments, the
membrane {\it must} be involved in order to define a sense
of rotation.   The speculative ways this could happen may
be classified according to the rules of microtubule
growth mentioned in the preceding subsection.
(a). Perhaps the microtubule-associated protein that 
nucleates a branch is also membrane associated [see Rule (ii), above].
(b) Perhaps the outcome of an mt-mt collision depends on which 
side the growing mt is impinging from [Rule (iv)], 
conceivably by the growing tip following a helical path
of the filaments and getting pushed into or away from the 
membrane, but just as likely by another membrane/microtubule
associated protein.

Imagine the mt-rotation mechanism (b) [the one based on collisions]
in more detail.  If new fibers have an orientation such that they 
hit old fibers from the right, then they grow longer. 
Thus, the new fibers' orientation tends to be rotated
{\it counter-clockwise} relative to the old fibers.

\OMIT{Reversal mutants?
Perhaps different mechanisms are simultaneously operating, thus
$\omega=\omega_1 - \omega_2$.
If we knock out $\omega_1$ $\Rightarrow$ in a mutation, then 
$\omega$ change sign, but it is not a literal reversal of the
whole mechanism.}

\section{Discussion and conclusions}

In conclusion, I reiterate: {\it symmetry is key} to recognizing
which mechanisms can possibly be responsible for L/R asymmetry.
Any such mechanism must explicitly connect three ingredients:
two axes, and some chiral molecule that implements the ``right
hand rule''.  Every story has a cell-level half and an organism
half. At the cell level, a natural axis is the membrane normal,
thus most of these mechanisms involved aligned ``cortical'' 
(adjacent to membrane) arrays of fibers.  

I conjecture that the mechanism is always cytoskeletal,
involving microtubules or actin fibers.
As for exactly how chirality enters:
the most elegant ``physics'' answer would be
the fiber itself, via some {\it screw mechanism} 
whereby motion along a long helical fiber
--- walking by a motor, or microtubule collisions~\footnote{
The stories in this paper omit a third kind of screw mechanism,
namely the change in pitch of a fiber (e.g. actin) under a change 
of its strains, or in its chemical environment.
See e.g. \cite{upmanyu}.}
gets converted into rotation around the axis.  However, a
``biology'' answer, a molecule anchored in the membrane and 
binding the fibers too, may be more plausible.  
The membrane-anchoring mechanism furnishes a hint to biologists 
as to which proteins
to focus on, in genetic or protein-expression studies aiming
to discover the master L/R determining gene.  
On the other hand, the screw-motion hypothesis suggests that, if a 
processive motor is involved, the mutations which 
reverse (or affect) L/R determination were those that
reversed (or affected) the motor's screw motion on its fiber.

Twice we were led into self-organized cortical arrays of
approximately parallel fibers  confined to the plane
adjacent to a membrane.  In both cases, the conjectured cell-level
mechanism did {\it not} depend on a globally defined $\xaxis$ axis, nor
did it even need an ordering of the fibers' polarization axes.
Instead, it used the orientation axis to define a {\it relative}
rotation, so the asymmetry was manifested in a rotation or shear rate. 

The brain asymmetry, which supplies the very terminology 
(``handedness'') for this subject, is the most mysterious.
Since nerve cells migrate far in the developing brain, 
one would hunt for an L/R asymmetry in cell locomotion.
But that would occur at a later stage,
in a more three-dimensional embryo, so it is less clear
what the locally defined $\xaxis$ and $\zaxis$ could be.
One possibility is that brain asymmetry is actually
a very early mechanism, perhaps the same one discussed
in Sec.~\ref{sec:celldiv}, as suggested by some
left-right anomalies seen in twins~\cite{levin-birthdefects}

It is left for future research to turn all these ideas into
quantitative estimates, a necessary condition before any
physics may be considered as completed.

\begin{theacknowledgments}
I thank Sourish Basu, Igor Segota, Michael Levin, 
Michael E. Fisher, and Eric D. Siggia  for discussions.
This work was supported by the U.S. Dept. of
Energy, grant DE-FG-ER45405.
\end{theacknowledgments}


\end{document}